\documentclass[submission,copyright,creativecommons]{eptcs}
\providecommand{\event}{HCVS 2025} 

\usepackage{iftex}

\ifpdf
  \usepackage{underscore}         
  \usepackage[T1]{fontenc}        
\else
  \usepackage{breakurl}           
\fi

\usepackage{graphicx}
\usepackage{orcidlink}
\usepackage{amsmath,amssymb}
\usepackage{listings}
\usepackage{tikz}
\usepackage{multirow}
\usepackage{collcell}
\usepackage{booktabs}
\usepackage{float}
\usepackage{enumitem}
\usepackage{rotating}
\usepackage{subcaption}
\usepackage{numprint}
\npthousandsep{\,}
\usepackage{array}
\usepackage{float}
\newcommand{\PreserveBackslash}[1]{\let\temp=\\#1\let\\=\temp}
\newcolumntype{C}[1]{>{\PreserveBackslash\centering}p{#1}}
\newcolumntype{R}[1]{>{\PreserveBackslash\raggedleft}p{#1}}
\newcolumntype{L}[1]{>{\PreserveBackslash\raggedright}p{#1}}

\usetikzlibrary{shapes.geometric}
\usetikzlibrary{positioning}
\usetikzlibrary{arrows}
\usetikzlibrary{shapes.multipart}
\usetikzlibrary{fit,calc}
\usepackage{pgfplots}
\usepackage{pgfplotstable}
\pgfplotsset{compat=1.18}

\makeatletter
\let\orgdescriptionlabel\descriptionlabel
\renewcommand*{\descriptionlabel}[1]{%
  \let\orglabel\label
  \let\label\@gobble
  \phantomsection
  \edef\@currentlabel{#1\unskip}%
  \let\label\orglabel
  \orgdescriptionlabel{#1}%
}
\makeatother

\title{\toolname{}: \\A Portfolio-Based Solver for Constrained Horn Clauses}

\author{Mihály Dobos-Kovács\orcidlink{0000-0002-0064-2965} \hspace{5ex} Levente Bajczi\orcidlink{0000-0002-6551-5860} \hspace{5ex} András Vörös\orcidlink{0000-0001-7617-3563}
\institute{Department of Artificial Intelligence and Systems Engineering\\Budapest University of Technology and Economics, Hungary\\\email{\{mdobosko,bajczi,vori\}@mit.bme.hu}}}

\newcommand{\titlerunning}{\textsc{CHCVerif}}
\newcommand{\authorrunning}{Dobos-Kovács et al.}

\hypersetup{
  bookmarksnumbered,
  pdftitle    = {\titlerunning},
  pdfauthor   = {\authorrunning},
  pdfsubject  = {HCVS'25},
  pdfkeywords = {CHC, SV-COMP, portfolio} 
}

\definecolor{ftsrg@AccentBlue}{RGB}{20,70,160}
\definecolor{ftsrg@AccentRed}{RGB}{150,0,24}
\definecolor{ftsrg@AccentPurple}{RGB}{82,43,71}
\definecolor{ftsrg@AccentOrange}{RGB}{251,139,36}
\definecolor{ftsrg@AccentLightBlue}{RGB}{68,114,196}
\definecolor{ftsrg@AccentGreen}{RGB}{112,173,71}

\newcommand{\toolname}{\textsc{CHCVerif}}
\newcommand{\numbercircle}[1]{%
  \begin{tikzpicture}[baseline=(char.base)]%
    \node[draw,circle,inner sep=1pt](char) {#1};%
  \end{tikzpicture}%
}

\begin{document}

\maketitle

\begin{abstract}
Constrained Horn Clauses (CHCs) are widely adopted as intermediate representations for a variety of verification tasks, including safety checking, invariant synthesis, and interprocedural analysis. This paper introduces \toolname{}, a portfolio-based CHC solver that adopts a software verification approach for solving CHCs. This approach enables us to reuse mature software verification tools to tackle CHC benchmarks, particularly those involving bitvectors and low-level semantics. Our evaluation shows that while the method enjoys only moderate success with linear integer arithmetic, it achieves modest success on bitvector benchmarks. Moreover, our results demonstrate the viability and potential of using software verification tools as backends for CHC solving, particularly when supported by a carefully constructed portfolio.
\end{abstract}

\begin{footnotesize}
		\paragraph{\footnotesize{Funding.}}
		This research was partially funded by the 2024-2.1.1-EKOP-2024-00003 University Research Scholarship Programme under project numbers EKOP-24-3-BME-213, and the Doctoral Excellence Fellowship Programme under project numbers DKÖP-23-1-BME-5 and DKÖP-23-1-BME-15; with the support provided by the Ministry of Culture and Innovation of Hungary from the NRDI Fund.

\end{footnotesize}

\section{Introduction}
\label{introduction}

Constrained Horn Clauses (CHCs) form a widely adopted intermediate representation for a variety of verification tasks, including safety checking \cite{Wesley2022, esen2022tricera, gurfinkel2015seahorn}, invariant synthesis \cite{Hu2020, Kim2021, Fedyukovich2019}, and systems verification \cite{Daniel2016}. Consequently, CHC solving has become a fundamental component in many formal verification pipelines. Despite the maturity of existing solvers such as Eldarica \cite{hojjat2018eldarica} and Spacer \cite{komuravelli2016smt}, there are still unsolved challenges, such as supporting CHCs with bitvector theories.

We present \toolname{}, a portfolio-based CHC solver that adopts a software verification perspective. Rather than solving CHCs directly at the logical level, \toolname{} translates CHC problems into semantically equivalent C programs and applies a portfolio of C verifiers to determine their correctness. The satisfiability of the original CHC system corresponds to the correctness of the generated C program. This approach enables \toolname{} to take advantage of mature software verification backends while diversifying its solving strategy through portfolio composition.

\toolname{} builds on \textsc{CoVeriTeam} \cite{beyer2022coveriteam}, a recently proposed framework for constructing compositional verification portfolios. \textsc{CoVeriTeam} allows \toolname{} to orchestrate multiple C verifiers within a configurable and extensible architecture. By expressing CHC solving as a software verification task and managing diverse tools through \textsc{CoVeriTeam}, \toolname{} offers a novel and modular pathway to Horn clause verification.

The remainder of the paper is structured as follows. Section \ref{sec:background} provides background on portfolio-based verification. Section \ref{sec:verification-approach} describes our overall approach to verification. Section \ref{sec:portfolio-construction} presents our process of constructing the portfolio. In Section \ref{sec:discussion}, we reflect on our results.

\section{Background}
\label{sec:background}

\subsection{CHC to C transformation}

Let us define a CHC problem as a set of \emph{deduction rules} in the context of this work. A deduction rule consists of a premise and a consequence. The premise can have zero or more \emph{uninterpreted functions}, while the consequence can have zero or exactly one.

If there are no uninterpreted functions in the premise, we call that rule an \emph{atom}. If there is more than one uninterpreted function in the premise of a rule, we call that rule \emph{non-linear}. If a deduction rule deduces the literal \emph{false}, we call that rule a \emph{query}.

If any rule is non-linear in a CHC problem, we call that problem \emph{non-linear}. Otherwise, we say that the CHC is \emph{linear}. 

In practical applications, CHC problems are frequently represented using the SMT-LIBv2 format~\cite{smtlibv2,DeAngelis2022}. This format depicts deduction rules as \emph{imply} expressions over a specific SMT theory. Moreover, every variable involved in the deduction rules is subject to universal quantification across the entire domain of variables, as required by the respective SMT theories. 

In this paper, we focus on the SMT theories \emph{core}, \emph{linear integer arithmetic}, and \emph{fixed-size bitvectors}. Note that problems requiring support for more theories exist (such as those using \emph{arrays} or \emph{algebraic data types}), but we discount those in the context of this work.

In the following, we present a simple example to showcase the basic idea behind the CHC to C transformation from \cite{somorjai2024bottoms, bajczi2025solving}.

\begin{align*}
A(x) \leftarrow x = 1 \\
A(x) \leftarrow A(x-1) \\
\mathit{false} \leftarrow A(11)
\end{align*}

The first rule declares that $A(1)$ is \emph{true}. The second rule propagates this fact forward, stating that if $A(x)$ holds, then $A(x+1)$ must also hold. Finally, the third rule is a \emph{query} that leads to a contradiction if $A(11)$ holds. Intuitively, this system is unsatisfiable: by repeated application of the second rule, starting from $A(1)$, we deduce $A(2)$, $A(3)$, and so on up to $A(11)$. Thus, $A(11)$ is deduced, which triggers the contradiction, making the system unsatisfiable.

The CHC system can be encoded in software in a top-down or bottom-up manner \cite{somorjai2024bottoms}. In the backward (top-down) encoding in \autoref{lst:backward-b}, the verifier starts with the query $A(11)$ and recursively unfolds the rules until it reaches the exit condition. Note that this program might fall into an infinite recursion based on the value of the parameter of the function $A$. However, tools that can reason about recursion may find that the exit condition of $A(11)$ is reachable.

In contrast, the forward (bottom-up) version shown in \autoref{lst:forward-b} constructs the program starting from the facts. It iteratively applies the rules using a loop and a nondeterministic choice of the current state. By explicitly modeling the inference steps, this version avoids recursion and can often be more amenable to automated reasoning by tools that support nondeterminism and state exploration, but in return, it only works with linear CHCs.

\begin{figure}[t]
\centering
\begin{minipage}[t]{0.4\textwidth}\centering\scriptsize
\begin{lstlisting}[numbers=left,escapechar=\%,language=C,caption={Backward transformation},label={lst:backward-b}]
int A(int x) {
    if(x == 1) return 1;
    else if(A(x - 1)) return 1;
    else return 0;
}

int main() {
    if(A(11)) return -1;
    else return 0;
}
\end{lstlisting}
\end{minipage}%
\begin{minipage}[t]{0.4\textwidth}\centering\scriptsize
\begin{lstlisting}[numbers=left,escapechar=\%,language=C,caption={Forward transformation},label={lst:forward-b}]
int main() {
    int A = 1, x;

    while(true) {
        x = nondet();
        if(A == 11) return -1;
        else if(A == x - 1) A = x;
    }
}
\end{lstlisting}
\end{minipage}%
\end{figure}

\subsection{Algorithm selection and portfolios}

Formal verification, in general, is a resource-intensive task that involves a variety of analysis techniques, such as symbolic execution, bounded model checking, abstract interpretation, and others, each with its own strengths and limitations. Given the inherent variability in how different tools perform across verification problems, using portfolio-based and algorithm selection strategies has received increasing attention in recent years \cite{beyer2025improvements}.

In the context of formal verification, the task of algorithm selection involves determining the most suitable algorithm, configuration, or tool from a set of options to solve a verification problem. The choice is typically based on characteristics derived from the problem, such as its structural details, control flow, data types, or recognizable patterns.

In contrast, a portfolio involves the execution of multiple verification tools or configurations on the same verification problem, either in parallel or in a coordinated sequence. The main idea is to make use of the complementary strengths of different verifiers, as no single tool consistently outperforms all others across the diverse landscape of verification tasks. While portfolios can combine configurations of a single tool, usually they are used to integrate tools using fundamentally different verification approaches (e.g., bounded model checking, abstract interpretation, or interpolation), improving overall robustness and increasing the likelihood of verification success within given time or resource bounds.

In recent years, the use of portfolio-based algorithm selection and techniques has grown significantly in the field of software verification \cite{beyer2025improvements}. Clear evidence of this fact is that approximately a third of the entries in SV-COMP 2025 \cite{beyer2025improvements} used some form of algorithm selection or portfolio in their verification approach. Entries include \textsc{Theta} \cite{theta-fmcad2017} or \textsc{CPAChecker} \cite{beyer2011cpachecker}, for example, that use select different algorithms based on the input problem, or CPV \cite{CPV-SVCOMP24} that uses a portfolio of other tools for verification.

\subsection{\textsc{CoVeriTeam}}

\textsc{CoVeriTeam} \cite{beyer2022coveriteam} is a framework for constructing and executing modular verification workflows by composing existing verification tools into customizable pipelines. It builds on three basic concepts: verification artifacts, actors, and compositional operators.

\begin{enumerate}
    \item Verification artifacts depict the data in the workflow, such as the input files, the property to verify, the results, and witnesses.
    \item Actors, on the other hand, act on the data and execute tools such as verifiers, validators, or test generators. They take verification artifacts as inputs and produce verification artifacts as outputs.
    \item Compositional operators provide a way to tie multiple actors together by taking care of scheduling tasks and directing the verification artifacts between the actors.
\end{enumerate}

\textsc{CoVeriTeam} is tightly integrated with the toolchain of SV-COMP. It executes the actors by using \textsc{BenchExec} \cite{beyer2019reliable} under the hood for reliable containerization and measurement. Moreover, as part of the submission process to SV-COMP, tool authors must submit a tool entry file to the FM Tools repository\footnote{\url{https://gitlab.com/sosy-lab/benchmarking/fm-tools}} that \textsc{CoVeriTeam} can consume as an actor definition. The tool entry file contains, among other information:

\begin{itemize}
    \item contains details on where to download the tool from. It supports DOIs for Zenodo archives or URLs for direct repository artifacts;
    \item describes which command-line parameters to pass to the tool;
    \item determines which \textsc{BenchExec} tool-info module to use that integrates the tool with the benchmark environment;
    \item lists dependencies that the system must have for the tool to run.
\end{itemize}

Leveraging this tight integration, \textsc{CoVeriTeam} is able to execute any tool from the last couple of editions of SV-COMP out-of-the-box, and makes it easy and well documented to integrate additional tools with it.

\section{Verification Approach}
\label{sec:verification-approach}

\begin{figure}[t]
    \centering
    \includegraphics[width=0.7\textwidth]{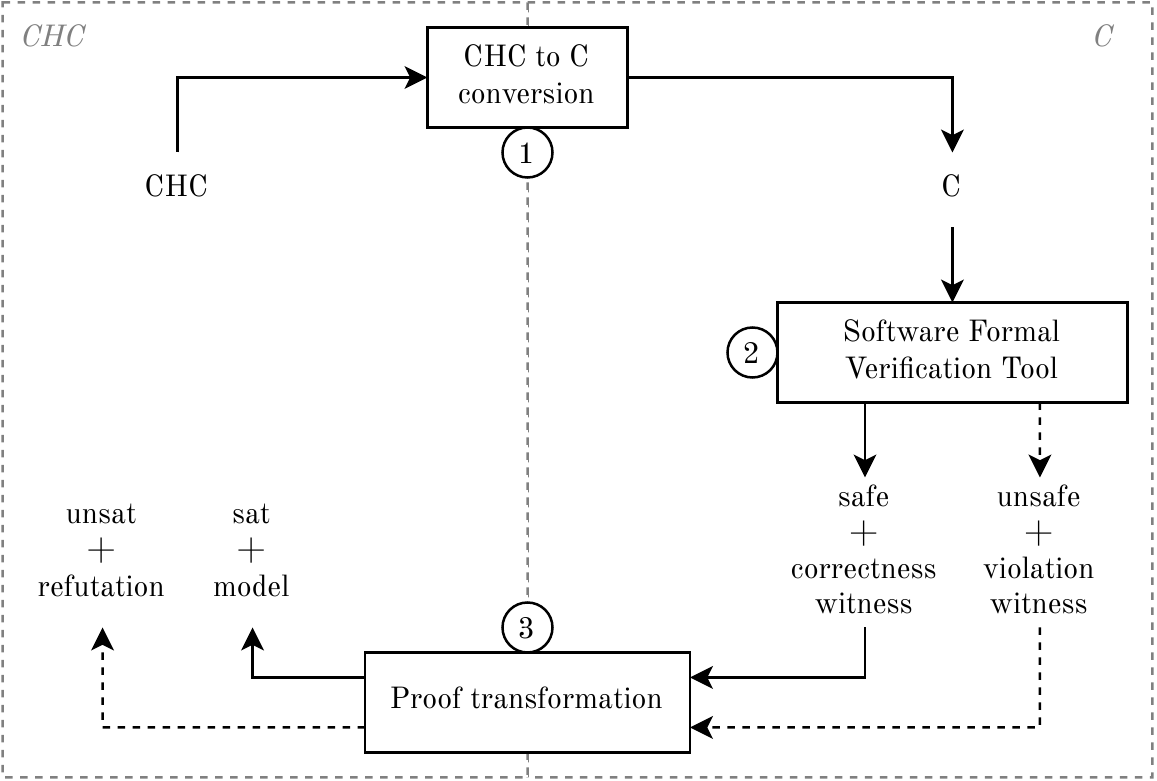}
    \caption{Overview of the proposed approach}
    \label{fig:overview}
\end{figure}

In this section, we present our approach to the verification of CHCs. Our main idea is based on our previous work in \cite{somorjai2024bottoms, bajczi2025solving} in which we presented a novel way to verify CHCs by transforming them into a C program (\numbercircle{1}). This transformation is done in such a way that the reachability of an assertion failure in the C program corresponds to the satisfiability of the CHC. The C program is then verified using a software verification tool (\numbercircle{2}) that produces a correctness witness if the program is safe (assertion failure is not reachable), or a violation witness if the program is unsafe (assertion failure is reachable). Finally, the verdicts of the software verification tool are transformed into the CHC domain (\numbercircle{3}): a safe verdict implies a satisfiable CHC, and the model can be extracted from the correctness witness, while an unsafe verdict means an unsatisfiable CHC, and the refutation can be derived from the violation witness. The approach is summarized in Figure \ref{fig:overview}.

In this paper, we focus only on \numbercircle{2}. More specifically, we aim to determine an optimal portfolio of software verification tools for C programs produced from CHCs. With this approach, we aim to diversify the field of CHC solving further, especially for theories where tool support is sparse (e.g., bitvectors).

\section{Portfolio construction}
\label{sec:portfolio-construction}

This section elaborates on the main ideas and the experiments conducted to determine the optimal portfolio. First, we discuss CHCs containing linear integer arithmetic and elaborate on how we overcame the challenges posed by the C standard for this category. Then, we explain our portfolio approach and present the results supporting our portfolios for CHCs containing bitvercors.

\subsection{Linear integer arithmetic}

The theory of linear integer arithmetic (LIA) deals with formulas over integer variables using linear expressions and supporting operations such as addition, subtraction, and comparisons, but excludes the multiplication of variables or non-linear terms.

One of the key differences between CHCs and C programs in handling integers is the way in which they are interpreted. CHCs use unbounded integers as they reason over the mathematical set of integers. C programs, on the other hand, are expected to run on hardware with finite resources. Thus, every value (in memory) is expressed over a finite number of bits, leading to a bounded domain.

This difference in semantics means that not all CHCs containing linear integer arithmetic can be mapped to a C program with equivalent behavior. Some CHCs might require the corresponding C program to evaluate expressions to values outside the possible domain; in other words, overflow. If the overflowing value is of an unsigned type, the value wraps around the domain, which does not match the behavior of such values in CHCs. On the other hand, if the overflowing value is of a signed type, the resulting behavior is undefined according to the C standard. Due to overflow, our approach can unfortunately be both unsound (with integer wraparounds causing infeasible traces) and incomplete (with a constrained integer domain causing incomplete models). 

\begin{figure}[t]
    \centering
    \includegraphics[width=0.7\textwidth]{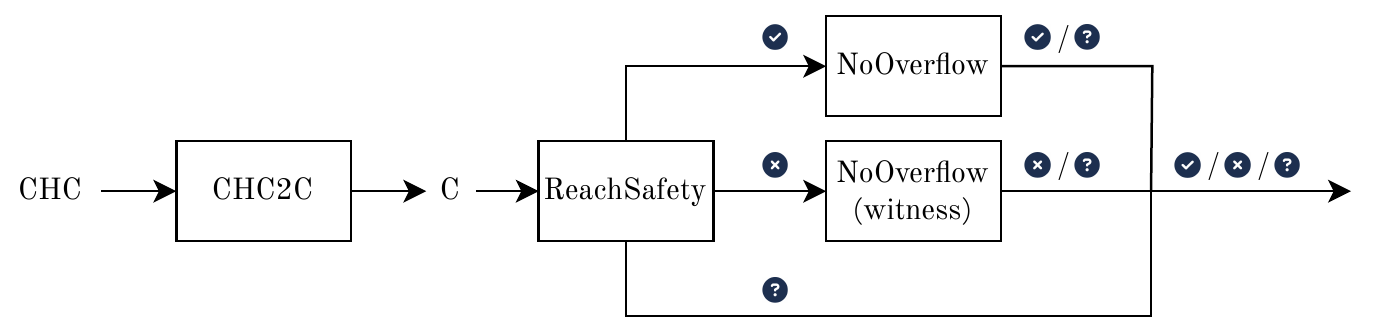}
    \caption{The proposed portfolio for the LIA theory}
    \label{fig:lia-portfolio1}
\end{figure}

Fortunately, there are software verification tools that can determine whether a C program contains an overflow or not. Based on this observation, we propose the following structure for the portfolio (Figure \ref{fig:lia-portfolio1}). We first transform the CHC into a C program via the transformation introduced in \cite{bajczi2025solving}. Then, we run the reachability analysis on the C program that can produce three different verdicts:

\begin{enumerate}
    \item If the verdict of the reachability analysis is safe, we check via an overflow analysis if the C program contains an overflow. If it does, the result is unsound, and the final verdict must be unknown. If there is no overflow in the problem, then we return a safe (satisfiable) result.
    \item If the verdict of the reachability analysis is unsafe, we generate a test case from the violation witness. If the execution of the witness leads to an overflow, the result is unsound, and the final verdict is unknown. If there is no overflow, the witness corresponds to a sound refutation, and we return an unsafe (unsatisfiable) result.
    \item If the verdict of the reachability analysis is unknown, we return that unknown verdict.
\end{enumerate}

To determine which tools to use, we opted to run experiments. We transformed approximately 600 CHCs from the benchmark set\footnote{\url{https://github.com/chc-comp/chc-comp24-benchmarks}} of CHC-COMP 2025 into C programs with both the recursive backward and the nonrecursive forward transformation. In the end, we ended up with 557 recursive C programs from the backward transformation of all CHCs and 195 nonrecursive C programs from the forward transformation of linear CHCs. We conducted our experiments using \textsc{BenchExec} \cite{beyer2019reliable}.

We selected 16 different tools based on the results of the NoOverflow category of SV-COMP 2025 \cite{beyer2025improvements}. We ran each benchmark with a time limit of 15 minutes, on 4 CPU cores and a memory limit of 15 GB of RAM. The results can be seen in Table \ref{tab:overflowresults}. The first column denotes whether the C program resulted from the nonrecursive forward transformation or the recursive backward transformation. The verdict column describes the verdict given by the overflow checker, while the verdict type describes whether the given verdict was correct or not. The rest of the columns contain the number of overflow tasks solved with the given resources.

The only tool that produced wrong results was \textsc{SVF-SVC}, but it produced more incorrect verdicts than correct ones, so it is less than an ideal candidate tool for verification. There were also tools (e.g., \textsc{Goblint}, \textsc{Theta}) that were able to prove if a program did not contain overflow, but failed to find an instance of overflow during the experiment, also rendering them unsuitable candidates for this task. Taking also into account the tool diversity as well (\textsc{UKojak}, \textsc{UTaipan}, and \textsc{UAutomizer} are the same tool family), we opted to use \textsc{Bubaak}, \textsc{Symbiotic}, \textsc{UAutomizer}, and \textsc{ESBMC-KIND} for overflow analysis.

\begin{table}[ht]
    \centering
    \caption{The number of solved overflow tasks with LIA}
    \label{tab:overflowresults}
    \scriptsize
    \renewcommand{\arraystretch}{1.1}
    \setlength{\tabcolsep}{3pt}\scriptsize
    \begin{tabular}{lllrrrrrrrrrrrrrrrrr}
        \toprule
        \textbf{Recursive} & \textbf{Verdict} & \textbf{Verdict type} &
        \rotatebox{90}{goblint} & 
        \rotatebox{90}{mopsa} & 
        \rotatebox{90}{emergentheta} & 
        \rotatebox{90}{sv-sanitizers} & 
        \rotatebox{90}{thorn} & 
        \rotatebox{90}{theta} & 
        \rotatebox{90}{svf-svc} & 
        \rotatebox{90}{bubaak-split} & 
        \rotatebox{90}{cpachecker} & 
        \rotatebox{90}{2ls} & 
        \rotatebox{90}{ukojak} & 
        \rotatebox{90}{bubaak} & 
        \rotatebox{90}{utaipan} & 
        \rotatebox{90}{symbiotic} & 
        \rotatebox{90}{uautomizer} & 
        \rotatebox{90}{esbmc-kind} & 
        \textbf{Out of} \\
        \midrule
        \multirow[c]{3}{*}{False} & No overflow & confirmed & 26 & 24 & 22 & 9  & 22 & 22 & 81  & 21 & 27 & 28 & 32 & 21 & 33 & 21 & 36 & 0 & \multirow[c]{3}{*}{195} \\ \cmidrule{2-19}
         & \multirow[c]{2}{*}{Overflow} & confirmed &  0 &  0 &  0 & 0  &  0 &  0 &  0  & 61 & 65 & 75 & 86 & 92 & 93 & 93 & 96 &105 & \\
         &  & wrong &  0 &  0 &  0 & 0  &  0 &  0 & 105 &  0 &  0 &  0 &  0 &  0 &  0 &  0 &  0 & 0 & \\ \midrule
        \multirow[c]{3}{*}{True} & No overflow & confirmed & 45  & 28  & 43  & 43  & 43  & 43  & 224 & 28  & 65  & 72  & 79  & 29  & 76  & 66  & 93  & 88  & \multirow[c]{3}{*}{557}\\ \cmidrule{2-19}
         & \multirow[c]{2}{*}{Overflow} & confirmed & 0   & 0   & 0   & 0   & 0   & 0   & 0   & 9   & 68  & 177 & 217 & 263 & 276 & 280 & 288 & 289 & \\
         &  & wrong & 0   & 0   & 0   & 0   & 0   & 0   & 304 & 0   & 0   & 0   & 0   & 0   & 0   & 0   & 0   & 0 &   \\

        \bottomrule
    \end{tabular}
\end{table}

For the reachability analysis, we selected 20 tools based on the results of the ReachSafety category of SV-COMP 2025. We ran each benchmark with a time limit of 15 minutes, on 4 CPU cores and a memory limit of 15 GB of RAM, similarly to the overflow experiments. The results can be seen in Table \ref{tab:reachabilityresults}. The first column denotes whether the C program resulted from the forward or backward transformation. The verdict column describes the verdict given by the reachability checker, while the verdict type describes whether the given verdict was correct or not. An unconfirmed verdict means that the correctness of the verdict is unknown, as it was neither confirmed nor refuted by any of the CHC solvers. The remaining columns contain the number of reachability tasks solved with the given resources.

The table highlights a wide variance in tool performance. For nonrecursive tasks, tools like \textsc{Bubaak} \cite{BUBAAK-SVCOMP25} and \textsc{Bubaak-split} \cite{BUBAAKSPLIT-SVCOMP24} show strong results with many confirmed safe verdicts (115 and 118 tasks, respectively). Taking unsafe verdicts into account as well, \textsc{Thorn} \cite{THETA-SVCOMP25} and \textsc{UAutomizer} \cite{ULTIMATE-SVCOMP13} rise as suitable candidates. Conversely, for recursive tasks, tools \textsc{CPV} \cite{CPV-SVCOMP24}, \textsc{UKojak} \cite{KOJAK-SVCOMP15}, and \textsc{UTaipan} \cite{UTAIPAN-SVCOMP23} lead the charge. The number of unconfirmed and wrong verdicts varies across tools and verdict types, indicating differences in precision and reliability. In the end, we opted to use \textsc{Thorn}, \textsc{Bubaak}, and \textsc{UTaipan} for the nonrecursive programs, while \textsc{CPV} and \textsc{UKojak} for the recursive programs.

Finally, we used \textsc{CPA-witness2test} \cite{beyer2018tests} to validate the unsafe result. Since converting violation witnesses to test cases and running them is not a resource-intensive task, and \textsc{CPA-witness2test} managed to handle every violation witness our tools produced, we opted not to explore other tools.

\begin{table}
    \centering
    \caption{The number of solved reachability tasks with LIA}
    \label{tab:reachabilityresults}
    \scriptsize
    \renewcommand{\arraystretch}{1.1}
    \setlength{\tabcolsep}{3pt}\scriptsize
    \begin{tabular}{lllrrrrrrrrrrrrrrrrrrrrr}
        \toprule
        \textbf{Recursive} & \textbf{Verdict} & \textbf{Verdict type} & 
        \rotatebox{90}{2ls} & 
        \rotatebox{90}{aise} & 
        \rotatebox{90}{brick} & 
        \rotatebox{90}{bubaak} & 
        \rotatebox{90}{bubaak-split} & 
        \rotatebox{90}{cpachecker} & 
        \rotatebox{90}{cpv} & 
        \rotatebox{90}{emergentheta} & 
        \rotatebox{90}{esbmc-kind} & 
        \rotatebox{90}{goblint} & 
        \rotatebox{90}{hornix} & 
        \rotatebox{90}{korn} & 
        \rotatebox{90}{mopsa} & 
        \rotatebox{90}{svf-svc} & 
        \rotatebox{90}{symbiotic} & 
        \rotatebox{90}{theta} & 
        \rotatebox{90}{thorn} & 
        \rotatebox{90}{uautomizer} & 
        \rotatebox{90}{ukojak} & 
        \rotatebox{90}{utaipan} &
        \textbf{Out of}\\
        \midrule
        \multirow[c]{6}{*}{False} & \multirow[c]{3}{*}{safe} & confirmed & 14 & 3 & 0 & 115 & 118 & 43 & 70 & 43 & 51 & 11 & 12 & 0 & 0 & 9 & 127 & 2 & 73 & 111 & 85 & 85 & \multirow[c]{6}{*}{195} \\
         &  & unconfirmed & 0 & 0 & 0 & 20 & 20 & 0 & 0 & 1 & 1 & 0 & 0 & 0 & 0 & 0 & 22 & 0 & 0 & 2 & 1 & 1 & \\
         &  & wrong & 0 & 0 & 0 & 18 & 22 & 0 & 0 & 0 & 1 & 0 & 0 & 0 & 0 & 0 & 37 & 0 & 4 & 5 & 4 & 2 & \\ \cmidrule{2-23}
         & \multirow[c]{3}{*}{unsafe}  & confirmed & 15 & 24 & 0 & 19 & 7 & 25 & 26 & 13 & 21 & 25 & 0 & 0 & 0 & 0 & 0 & 20 & 24 & 28 & 27 & 27 & \\
         &  & unconfirmed & 1 & 2 & 0 & 2 & 2 & 4 & 1 & 2 & 0 & 2 & 0 & 0 & 0 & 0 & 0 & 2 & 1 & 0 & 0 & 0 & \\
         &  & wrong & 13 & 14 & 0 & 12 & 9 & 14 & 6 & 13 & 0 & 18 & 0 & 0 & 0 & 0 & 0 & 14 & 0 & 0 & 0 & 0 & \\ \midrule
        \multirow[c]{6}{*}{True} & \multirow[c]{3}{*}{safe}  & confirmed & 21 & 6 & 0 & 24 & 25 & 60 & 96 & 19 & 18 & 21 & 0 & 13 & 13 & 12 & 336 & 23 & 24 & 19 & 132 & 132 & \multirow[c]{6}{*}{557} \\
         &  & unconfirmed & 0 & 0 & 0 & 0 & 0 & 0 & 0 & 0 & 0 & 0 & 0 & 0 & 0 & 0 & 50 & 0 & 0 & 0 & 0 & 0 & \\
         &  & wrong & 0 & 0 & 0 & 0 & 0 & 2 & 3 & 0 & 0 & 0 & 0 & 0 & 0 & 0 & 142 & 0 & 1 & 0 & 0 & 0 & \\ \cmidrule{2-23}
         & \multirow[c]{3}{*}{unsafe} & confirmed & 13 & 2 & 0 & 80 & 41 & 61 & 69 & 8 & 11 & 110 & 0 & 7 & 1 & 0 & 0 & 84 & 24 & 12 & 107 & 101 & \\
         &  & unconfirmed & 0 & 0 & 0 & 2 & 2 & 1 & 0 & 0 & 0 & 2 & 0 & 0 & 0 & 0 & 0 & 2 & 1 & 0 & 0 & 0 & \\
         &  & wrong & 3 & 0 & 0 & 20 & 16 & 16 & 1 & 3 & 0 & 24 & 0 & 0 & 0 & 0 & 0 & 20 & 1 & 0 & 0 & 0 & \\
        
        \bottomrule
    \end{tabular}
\end{table}

In comparison, we included the result of some CHC solvers on the same benchmark set in Table \ref{tab:chc-lia}. It can be seen that while software verification tools tend to solve around 100 CHCs, the CHC solvers are in the magnitude of 300-400.

\begin{table}
    \centering
    \caption{The number of solved CHC problems by CHC solvers for the LIA theory}
    \label{tab:chc-lia}
    \scriptsize
    \renewcommand{\arraystretch}{1.1}
    \setlength{\tabcolsep}{3pt}\scriptsize
    \begin{tabular}{lccccc}
        \toprule
        \textbf{Verdict} &
        \rotatebox{90}{eldarica} & 
        \rotatebox{90}{golem} & 
        \rotatebox{90}{ultimateunihorn} & 
        \rotatebox{90}{spacer} &
        \textbf{Out of} \\
        \midrule
        sat & 286 & 256 & 149 & 337 & \multirow[c]{2}{*}{557} \\
        unsat & 151 & 164 & 125 & 160 & \\
        \bottomrule
    \end{tabular}
\end{table}

\subsection{Bitprecise arithmetic}

The verification of CHCs with linear integer arithmetic had to use a portfolio of both reachability and overflow analysis tools to bridge the gap between the semantics of mathematical integers and values in a C program. While CHCs with mathematical integers are prevalent in practice in the CHC-COMP benchmark suite, some CHCs use other theories, most notably bitvectors.

Bitvectors represent the values on a finite number of bits and, in 2's complement, capture the semantics of C values better than mathematical integers can. Unlike mathematical integers, bitvectors provide bit-level access to the value and wrap around when a value is out of their bounds.

While bitvectors are widely used, tool support in CHC solvers for bitvectors is sparse. Of the tools participating in CHC-COMP 2024, only \textsc{Theta} \cite{somorjai2024bottoms}, and \textsc{Eldarica} \cite{hojjat2018eldarica} were able to verify CHC problems with bitvectors. On the other hand, more than 25 tools participated in the ReachSafety-BitVectors category on SV-COMP 2024. Making use of these tools for CHC solving would significantly diversify solving capabilities.

As bitvectors capture the semantics of C values, there is no need to check for overflow or validate the result afterwards: the safety of the program implies the satisfiability of the CHC problem directly. Therefore, it is sufficient to perform a reachability analysis only for CHC problems with bitvectors.

We transformed the CHCs in the bitvectors category of the CHC-COMP 2024 benchmark suite\footnote{\url{https://github.com/chc-comp/chc-comp24-benchmarks}}, and ended up with 213 nonrecursive C programs and 396 recursive ones. We selected seven different tools based on the results of the ReachSafety-Bitvector category of SV-COMP 2025 \cite{beyer2025improvements}. We ran each benchmark with a time limit of 15 minutes, on 2 CPU cores and a memory limit of 15 GB of RAM. The results of the experiments are in Table \ref{tab:bv-results}.

Notably, in the nonrecursive setting, most tools achieve a relatively high number of confirmed unsafe verdicts, indicating strong effectiveness in proving unsatisfiability. Tools such as \textsc{CPAChecker} \cite{CPACHECKER-SVCOMP24}, \textsc{ESBMC-KIND} \cite{ESBMC-SVCOMP25}, and \textsc{Symbiotic} \cite{SYMBIOTIC-SVCOMP24} show consistently high confirmed unsat results. The number of confirmed safe verdicts is more modest and varies across tools (especially taking wrong verdicts into account), with \textsc{CPAChecker} achieving relatively higher counts. For recursive tasks, the landscape is similar, with the notable exception that only \textsc{CPAChecker} was able to produce more than a couple of safe verdicts.

\begin{table}
\centering
\caption{The number of solved reachability tasks for bitvectors }
\label{tab:bv-results}
\scriptsize
\renewcommand{\arraystretch}{1.1}
\setlength{\tabcolsep}{4pt}\scriptsize
\begin{tabular}{lllrrrrrrrr}
\toprule
\textbf{Recursive} & \textbf{Verdict} & \textbf{Verdict type} & 
\rotatebox{90}{2ls} & 
\rotatebox{90}{bubaak-split} & 
\rotatebox{90}{cpachecker} & 
\rotatebox{90}{emergentheta} & 
\rotatebox{90}{esbmc-kind} & 
\rotatebox{90}{symbiotic} & 
\rotatebox{90}{theta} &
\textbf{Out of}\\
\midrule
\multirow[c]{5}{*}{False} & \multirow[c]{3}{*}{safe}   & confirmed   & 9  & 39 & 27 & 16 & 5  & 0  & 38 & \multirow[c]{5}{*}{213}\\
      &       & unconfirmed & 4  & 28 & 6  & 21 & 4  & 0  & 32 & \\
      &       & wrong       & 0  & 16 & 0  & 4  & 0  & 0  & 8  & \\
          \cmidrule{2-10}
      &  \multirow[c]{2}{*}{unsafe} & confirmed   & 31 & 62 & 81 & 69 & 79 & 75 & 68 & \\
      &       & unconfirmed & 5  & 8  & 12 & 3  & 10 & 10 & 3 &  \\
\midrule
\multirow[c]{5}{*}{True}  & \multirow[c]{3}{*}{safe}   & confirmed   & 0  & 0  & 24 & 0  & 0  & 0  & 0 & \multirow[c]{5}{*}{396}  \\
      &       & unconfirmed & 1  & 1  & 13 & 5  & 1  & 1  & 6 &  \\
      &       & wrong       & 0  & 0  & 0  & 2  & 0  & 0  & 4 &  \\
          \cmidrule{2-10}
      & \multirow[c]{2}{*}{unsafe} & confirmed   & 69 & 88 & 91 & 67 & 101& 99 & 73 & \\
      &       & unconfirmed & 4  & 9  & 10 & 0  & 18 & 16 & 4 & \\
\bottomrule
\end{tabular}
\end{table}

We also included the performance of two CHC solvers on the same benchmark set in Table \ref{tab:chc-bv} for comparison. It can be seen that while the CHC solvers still perform better, the gap is far less than previously with linear integer arithmetic. Moreover, the relatively high ratio of unconfirmed to confirmed verdicts shows that the software tools solved numerous tasks that none of the CHC solvers were able to. In the end, we opted to use a portfolio of \textsc{CPAChecker}, \textsc{ESBMC-KIND}, and \textsc{Symbiotic}.

\begin{table}[ht]
    \centering
    \caption{The number of solved CHC problems by CHC solvers for the bitvector theory}
    \label{tab:chc-bv}
    \scriptsize
    \renewcommand{\arraystretch}{1.1}
    \setlength{\tabcolsep}{3pt}\scriptsize
    \begin{tabular}{lrrr}
        \toprule
        \textbf{Verdict} &
        \rotatebox{90}{eldarica} & 
        \rotatebox{90}{theta} &
        \textbf{Out of} \\
        \midrule
        sat & 103 & 44 & \multirow[c]{2}{*}{396} \\
        unsat & 160 & 161 & \\
        \bottomrule
    \end{tabular}
\end{table}

\subsection{Final portfolio}

Based on the previous findings, we opted to use the following portfolio at the end. We start by first transforming the CHC into a C program nonrecursively. After that, based on the theory used, the workflow diverges. If the CHC uses LIA, first, a reachability analysis is conducted with \textsc{Thorn}, \textsc{Bubaak}, and \textsc{UTaipan} in a parallel configuration (meaning the tools are running until one of them produces a result). Then, based on the verdict, the verdict is either validated via \textsc{CPA-witness2test} or by an overflow check with \textsc{Bubaak}, \textsc{Symbiotic}, \textsc{UAutomizer}, and \textsc{ESBMC-KIND} in a parallel configuration. If the CHC uses bitvectors, only a single reachability analysis is conducted by \textsc{CPAChecker}, \textsc{ESBMC-KIND}, and \textsc{Symbiotic} in a parallel configuration. If the analysis produces a result, the portfolio ends its run.

If the nonrecursive reachability or overflow analysis does not yield a result under a given time limit, the nonrecursive analysis is terminated. Next, the recursive backward transformation transforms the CHC into a C program. After that, the portfolio diverges again based on the theory used by the CHC. In the case of LIA, a reachability analysis by \textsc{CPV} and \textsc{UKojak} is followed by an overflow analysis with \textsc{Bubaak}, \textsc{Symbiotic}, \textsc{UAutomizer}, and \textsc{ESBMC-KIND} in a parallel configuration. If the CHC uses bitvectors, only a single reachability analysis is conducted by \textsc{CPAChecker}, \textsc{ESBMC-KIND}, and \textsc{Symbiotic}. 

We implemented the proposed portfolio in the open-source \toolname{} \cite{doboskovacs202515330957} tool using \textsc{CoVeriTeam} and using the archives of the aforementioned tools from SV-COMP 2024.

\section{Discussion}
\label{sec:discussion}

Our results demonstrate the viability and potential of using software verification tools as a backend for CHC solving, particularly when supported by a carefully constructed portfolio. This strategy opens up new directions for verifying CHCs over theories with limited or no dedicated solver support, such as bitvectors.

Bitvectors, in particular, represent a challenging theory for traditional CHC solvers due to their low-level, word-based semantics and their close correspondence with hardware behavior. However, bitvectors are widely supported by state-of-the-art software verifiers. By translating CHCs into C code, our approach effectively leverages the mature tool ecosystem of software verification to handle such cases, allowing us to shift the burden of reasoning about bit-level behavior onto tools that are already optimized for such tasks, such as \textsc{CPAChecker}, \textsc{ESBMC}, and \textsc{Symbiotic}.

In the case of the more established linear integer arithmetics, the overall picture is much more nuanced. First, the software verification tools performed better on the nonrecursive forward transformation that only supports linear CHCs, but are far from the performance of CHC solvers. The fact that the backward transformation performed worse is not surprising, as it introduces recursive functions into the verification process, which are considered a more difficult verification problem and are supported by fewer tools. However, the main limitation is the poor performance of the overflow tools. As these results need to be verified by an overflow check, the poor overflow analysis performance will impact the whole portfolio.

\section{Conclusion}
\label{sec:conclusion}

We presented \toolname{}, a tool for solving Constrained Horn Clause (CHC) problems by translating them to C programs and leveraging existing software verification tools to check safety properties. This approach enables the reuse of mature verification infrastructures to tackle CHC benchmarks, particularly those involving bitvectors and low-level semantics. Our evaluation shows that while the method enjoys only moderate success with linear integer arithmetic due to semantic mismatches, it achieves modest success on bitvector benchmarks, demonstrating the potential of this translation-based approach for certain classes of CHC problems. In the future, we plan to extend \toolname{} to support more theories, such as floating-point numbers that no CHC solvers support yet.

\bibliographystyle{eptcs}
\bibliography{chc,svcomp}

\begin{thebibliography}{10}
\providecommand{\bibitemdeclare}[2]{}
\providecommand{\surnamestart}{}
\providecommand{\surnameend}{}
\providecommand{\urlprefix}{Available at }
\providecommand{\url}[1]{\texttt{#1}}
\providecommand{\href}[2]{\texttt{#2}}
\providecommand{\urlalt}[2]{\href{#1}{#2}}
\providecommand{\doi}[1]{doi:\urlalt{https://doi.org/#1}{#1}}
\providecommand{\eprint}[1]{arXiv:\urlalt{https://arxiv.org/abs/#1}{#1}}
\providecommand{\bibinfo}[2]{#2}

\bibitemdeclare{inproceedings}{CPACHECKER-SVCOMP24}
\bibitem{CPACHECKER-SVCOMP24}
\bibinfo{author}{D.~\surnamestart Baier\surnameend},
  \bibinfo{author}{D.~\surnamestart Beyer\surnameend}, \bibinfo{author}{P.-C.
  \surnamestart Chien\surnameend}, \bibinfo{author}{M.~\surnamestart
  Jankola\surnameend}, \bibinfo{author}{M.~\surnamestart Kettl\surnameend},
  \bibinfo{author}{N.-Z. \surnamestart Lee\surnameend},
  \bibinfo{author}{T.~\surnamestart Lemberger\surnameend},
  \bibinfo{author}{M.~\surnamestart Lingsch-Rosenfeld\surnameend},
  \bibinfo{author}{M.~\surnamestart Spiessl\surnameend},
  \bibinfo{author}{H.~\surnamestart Wachowitz\surnameend} \&
  \bibinfo{author}{P.~\surnamestart Wendler\surnameend} (\bibinfo{year}{2024}):
  \emph{\bibinfo{title}{\textsc{CPAchecker} 2.3 with Strategy Selection
  (Competition Contribution)}}.
\newblock In: {\slshape \bibinfo{booktitle}{Proc.\ TACAS~(3)}},
  \bibinfo{series}{LNCS~14572}, \bibinfo{publisher}{Springer}, pp.
  \bibinfo{pages}{359--364}, \doi{10.1007/978-3-031-57256-2_21}.

\bibitemdeclare{inproceedings}{bajczi2025solving}
\bibitem{bajczi2025solving}
\bibinfo{author}{Levente \surnamestart Bajczi\surnameend} \&
  \bibinfo{author}{Vince \surnamestart Moln{\'a}r\surnameend}
  (\bibinfo{year}{2025}): \emph{\bibinfo{title}{Solving Constrained Horn
  Clauses as C Programs with CHC2C}}.
\newblock In \bibinfo{editor}{Thomas \surnamestart Neele\surnameend} \&
  \bibinfo{editor}{Anton \surnamestart Wijs\surnameend}, editors: {\slshape
  \bibinfo{booktitle}{Model Checking Software}}, \bibinfo{publisher}{Springer
  Nature Switzerland}, \bibinfo{address}{Cham}, pp. \bibinfo{pages}{146--163},
  \doi{10.1007/978-3-031-66149-5_8}.

\bibitemdeclare{inproceedings}{beyer2018tests}
\bibitem{beyer2018tests}
\bibinfo{author}{Dirk \surnamestart Beyer\surnameend},
  \bibinfo{author}{Matthias \surnamestart Dangl\surnameend},
  \bibinfo{author}{Thomas \surnamestart Lemberger\surnameend} \&
  \bibinfo{author}{Michael \surnamestart Tautschnig\surnameend}
  (\bibinfo{year}{2018}): \emph{\bibinfo{title}{Tests from Witnesses}}.
\newblock In \bibinfo{editor}{Catherine \surnamestart Dubois\surnameend} \&
  \bibinfo{editor}{Burkhart \surnamestart Wolff\surnameend}, editors: {\slshape
  \bibinfo{booktitle}{Tests and Proofs}}, \bibinfo{publisher}{Springer
  International Publishing}, \bibinfo{address}{Cham}, pp.
  \bibinfo{pages}{3--23}, \doi{10.1007/978-3-319-92994-1_1}.

\bibitemdeclare{inproceedings}{beyer2022coveriteam}
\bibitem{beyer2022coveriteam}
\bibinfo{author}{Dirk \surnamestart Beyer\surnameend} \&
  \bibinfo{author}{Sudeep \surnamestart Kanav\surnameend}
  (\bibinfo{year}{2022}): \emph{\bibinfo{title}{CoVeriTeam: On-demand
  composition of cooperative verification systems}}.
\newblock In: {\slshape \bibinfo{booktitle}{International Conference on Tools
  and Algorithms for the Construction and Analysis of Systems}},
  \bibinfo{organization}{Springer}, pp. \bibinfo{pages}{561--579},
  \doi{10.1007/978-3-030-99524-9\_31}.

\bibitemdeclare{inproceedings}{beyer2011cpachecker}
\bibitem{beyer2011cpachecker}
\bibinfo{author}{Dirk \surnamestart Beyer\surnameend} \&
  \bibinfo{author}{M~Erkan \surnamestart Keremoglu\surnameend}
  (\bibinfo{year}{2011}): \emph{\bibinfo{title}{CPAchecker: A tool for
  configurable software verification}}.
\newblock In: {\slshape \bibinfo{booktitle}{Computer Aided Verification: 23rd
  International Conference, CAV 2011, Snowbird, UT, USA, July 14-20, 2011.
  Proceedings 23}}, \bibinfo{organization}{Springer}, pp.
  \bibinfo{pages}{184--190}, \doi{10.1007/978-3-642-22110-1_16}.

\bibitemdeclare{article}{beyer2019reliable}
\bibitem{beyer2019reliable}
\bibinfo{author}{Dirk \surnamestart Beyer\surnameend}, \bibinfo{author}{Stefan
  \surnamestart L{\"o}we\surnameend} \& \bibinfo{author}{Philipp \surnamestart
  Wendler\surnameend} (\bibinfo{year}{2019}): \emph{\bibinfo{title}{Reliable
  benchmarking: requirements and solutions}}.
\newblock {\slshape \bibinfo{journal}{International Journal on Software Tools
  for Technology Transfer}} \bibinfo{volume}{21}(\bibinfo{number}{1}), pp.
  \bibinfo{pages}{1--29}, \doi{10.1007/s10009-017-0469-y}.

\bibitemdeclare{inproceedings}{beyer2025improvements}
\bibitem{beyer2025improvements}
\bibinfo{author}{Dirk \surnamestart Beyer\surnameend} \& \bibinfo{author}{Jan
  \surnamestart Strej{\v{c}}ek\surnameend} (\bibinfo{year}{2025}):
  \emph{\bibinfo{title}{Improvements in software verification and witness
  validation: SV-COMP 2025}}.
\newblock In: {\slshape \bibinfo{booktitle}{International Conference on Tools
  and Algorithms for the Construction and Analysis of Systems}},
  \bibinfo{organization}{Springer}, pp. \bibinfo{pages}{151--186},
  \doi{10.1007/978-3-031-90660-2_9}.

\bibitemdeclare{inproceedings}{BUBAAKSPLIT-SVCOMP24}
\bibitem{BUBAAKSPLIT-SVCOMP24}
\bibinfo{author}{M.~\surnamestart Chalupa\surnameend} \&
  \bibinfo{author}{C.~\surnamestart Richter\surnameend} (\bibinfo{year}{2024}):
  \emph{\bibinfo{title}{\textsc{Bubaak-SpLit}: {Split} What You Cannot Verify
  (Competition Contribution)}}.
\newblock In: {\slshape \bibinfo{booktitle}{Proc.\ TACAS~(3)}},
  \bibinfo{series}{LNCS~14572}, \bibinfo{publisher}{Springer}, pp.
  \bibinfo{pages}{353--358}, \doi{10.1007/978-3-031-57256-2_20}.

\bibitemdeclare{inproceedings}{BUBAAK-SVCOMP25}
\bibitem{BUBAAK-SVCOMP25}
\bibinfo{author}{M.~\surnamestart Chalupa\surnameend} \&
  \bibinfo{author}{C.~\surnamestart Richter\surnameend} (\bibinfo{year}{2025}):
  \emph{\bibinfo{title}{\textsc{Bubaak}: {Dynamic} Cooperative Verification
  (Competition Contribution)}}.
\newblock In: {\slshape \bibinfo{booktitle}{Proc.\ TACAS~(3)}},
  \bibinfo{series}{LNCS~15698}, \bibinfo{publisher}{Springer}, pp.
  \bibinfo{pages}{212--216}, \doi{10.1007/978-3-031-90660-2_14}.

\bibitemdeclare{inproceedings}{CPV-SVCOMP24}
\bibitem{CPV-SVCOMP24}
\bibinfo{author}{P.-C. \surnamestart Chien\surnameend} \&
  \bibinfo{author}{N.-Z. \surnamestart Lee\surnameend} (\bibinfo{year}{2024}):
  \emph{\bibinfo{title}{\textsc{CPV}: {A} Circuit-Based Program Verifier
  (Competition Contribution)}}.
\newblock In: {\slshape \bibinfo{booktitle}{Proc.\ TACAS~(3)}},
  \bibinfo{series}{LNCS~14572}, \bibinfo{publisher}{Springer}, pp.
  \bibinfo{pages}{365--370}, \doi{10.1007/978-3-031-57256-2_22}.

\bibitemdeclare{misc}{smtlibv2}
\bibitem{smtlibv2}
\bibinfo{author}{David~R. \surnamestart Cok\surnameend} (\bibinfo{year}{2012}):
  \emph{\bibinfo{title}{{The SMT-LIBv2 Language and Tools: A Tutorial}}}.
\newblock \urlprefix\url{https://smtlib.github.io/jSMTLIB/SMTLIBTutorial.pdf}.

\bibitemdeclare{inproceedings}{Daniel2016}
\bibitem{Daniel2016}
\bibinfo{author}{Jakub \surnamestart Daniel\surnameend},
  \bibinfo{author}{Alessandro \surnamestart Cimatti\surnameend},
  \bibinfo{author}{Alberto \surnamestart Griggio\surnameend},
  \bibinfo{author}{Stefano \surnamestart Tonetta\surnameend} \&
  \bibinfo{author}{Sergio \surnamestart Mover\surnameend}
  (\bibinfo{year}{2016}): \emph{\bibinfo{title}{{Infinite-State
  Liveness-to-Safety via Implicit Abstraction and Well-Founded Relations}}}.
\newblock In \bibinfo{editor}{Swarat \surnamestart Chaudhuri\surnameend} \&
  \bibinfo{editor}{Azadeh \surnamestart Farzan\surnameend}, editors: {\slshape
  \bibinfo{booktitle}{Computer Aided Verification - 28th International
  Conference, {CAV} 2016, Toronto, ON, Canada, July 17-23, 2016, Proceedings,
  Part {I}}}, {\slshape \bibinfo{series}{Lecture Notes in Computer Science}}
  \bibinfo{volume}{9779}, \bibinfo{publisher}{Springer}, pp.
  \bibinfo{pages}{271--291}, \doi{10.1007/978-3-319-41528-4\_15}.

\bibitemdeclare{inproceedings}{DeAngelis2022}
\bibitem{DeAngelis2022}
\bibinfo{author}{Emanuele \surnamestart {De Angelis}\surnameend} \&
  \bibinfo{author}{Hari Govind~V. \surnamestart K.\surnameend}
  (\bibinfo{year}{2022}): \emph{\bibinfo{title}{{CHC-COMP 2022: Competition
  Report}}}.
\newblock In \bibinfo{editor}{Geoffrey~William \surnamestart
  Hamilton\surnameend}, \bibinfo{editor}{Temesghen \surnamestart
  Kahsai\surnameend} \& \bibinfo{editor}{Maurizio \surnamestart
  Proietti\surnameend}, editors: {\slshape \bibinfo{booktitle}{Proceedings 9th
  Workshop on Horn Clauses for Verification and Synthesis and 10th
  International Workshop on Verification and Program Transformation,
  HCVS/VPT@ETAPS 2022, and 10th International Workshop on Verification and
  Program TransformationMunich, Germany, 3rd April 2022}}, {\slshape
  \bibinfo{series}{EPTCS}} \bibinfo{volume}{373}, pp. \bibinfo{pages}{44--62},
  \doi{10.4204/EPTCS.373.5}.

\bibitemdeclare{inproceedings}{UTAIPAN-SVCOMP23}
\bibitem{UTAIPAN-SVCOMP23}
\bibinfo{author}{D.~\surnamestart Dietsch\surnameend},
  \bibinfo{author}{M.~\surnamestart Heizmann\surnameend},
  \bibinfo{author}{D.~\surnamestart Klumpp\surnameend},
  \bibinfo{author}{F.~\surnamestart Schüssele\surnameend} \&
  \bibinfo{author}{A.~\surnamestart Podelski\surnameend}
  (\bibinfo{year}{2023}): \emph{\bibinfo{title}{\textsc{Ultimate Taipan} 2023
  (Competition Contribution)}}.
\newblock In: {\slshape \bibinfo{booktitle}{Proc.\ TACAS~(2)}},
  \bibinfo{series}{LNCS~13994}, \bibinfo{publisher}{Springer}, pp.
  \bibinfo{pages}{582--587}, \doi{10.1007/978-3-031-30820-8_40}.

\bibitemdeclare{}{doboskovacs202515330957}
\bibitem{doboskovacs202515330957}
\bibinfo{author}{Mihály \surnamestart Dobos-Kovács\surnameend} \&
  \bibinfo{author}{Levente \surnamestart Bajczi\surnameend}
  (\bibinfo{year}{2025}): \emph{\bibinfo{title}{chc2c-svcomp: Portfolio of
  software verification tools for solving systems of Horn clauses.}},
  \doi{10.5281/zenodo.15283157}.

\bibitemdeclare{inproceedings}{esen2022tricera}
\bibitem{esen2022tricera}
\bibinfo{author}{Zafer \surnamestart Esen\surnameend} \&
  \bibinfo{author}{Philipp \surnamestart Rümmer\surnameend}
  (\bibinfo{year}{2022}): \emph{\bibinfo{title}{Tricera: Verifying C Programs
  Using the Theory of Heaps}}.
\newblock In: {\slshape \bibinfo{booktitle}{2022 Formal Methods in
  Computer-Aided Design (FMCAD)}}, pp. \bibinfo{pages}{380--391},
  \doi{10.34727/2022/isbn.978-3-85448-053-2_45}.

\bibitemdeclare{inproceedings}{Fedyukovich2019}
\bibitem{Fedyukovich2019}
\bibinfo{author}{Grigory \surnamestart Fedyukovich\surnameend},
  \bibinfo{author}{Arie \surnamestart Gurfinkel\surnameend} \&
  \bibinfo{author}{Aarti \surnamestart Gupta\surnameend}
  (\bibinfo{year}{2019}): \emph{\bibinfo{title}{{Lazy but Effective Functional
  Synthesis}}}.
\newblock In \bibinfo{editor}{Constantin \surnamestart Enea\surnameend} \&
  \bibinfo{editor}{Ruzica \surnamestart Piskac\surnameend}, editors: {\slshape
  \bibinfo{booktitle}{Verification, Model Checking, and Abstract Interpretation
  - 20th International Conference, {VMCAI} 2019, Cascais, Portugal, January
  13-15, 2019, Proceedings}}, {\slshape \bibinfo{series}{Lecture Notes in
  Computer Science}} \bibinfo{volume}{11388}, \bibinfo{publisher}{Springer},
  pp. \bibinfo{pages}{92--113}, \doi{10.1007/978-3-030-11245-5\_5}.

\bibitemdeclare{inproceedings}{gurfinkel2015seahorn}
\bibitem{gurfinkel2015seahorn}
\bibinfo{author}{Arie \surnamestart Gurfinkel\surnameend},
  \bibinfo{author}{Temesghen \surnamestart Kahsai\surnameend},
  \bibinfo{author}{Anvesh \surnamestart Komuravelli\surnameend} \&
  \bibinfo{author}{Jorge~A \surnamestart Navas\surnameend}
  (\bibinfo{year}{2015}): \emph{\bibinfo{title}{The SeaHorn verification
  framework}}.
\newblock In: {\slshape \bibinfo{booktitle}{International Conference on
  Computer Aided Verification}}, \bibinfo{organization}{Springer}, pp.
  \bibinfo{pages}{343--361}, \doi{10.1007/978-3-319-21690-4_20}.

\bibitemdeclare{inproceedings}{ULTIMATE-SVCOMP13}
\bibitem{ULTIMATE-SVCOMP13}
\bibinfo{author}{M.~\surnamestart Heizmann\surnameend},
  \bibinfo{author}{J.~\surnamestart Christ\surnameend},
  \bibinfo{author}{D.~\surnamestart Dietsch\surnameend},
  \bibinfo{author}{E.~\surnamestart Ermis\surnameend},
  \bibinfo{author}{J.~\surnamestart Hoenicke\surnameend},
  \bibinfo{author}{M.~\surnamestart Lindenmann\surnameend},
  \bibinfo{author}{A.~\surnamestart Nutz\surnameend},
  \bibinfo{author}{C.~\surnamestart Schilling\surnameend} \&
  \bibinfo{author}{A.~\surnamestart Podelski\surnameend}
  (\bibinfo{year}{2013}): \emph{\bibinfo{title}{\textsc{Ultimate Automizer}
  with {SMTInterpol} (Competition Contribution)}}.
\newblock In: {\slshape \bibinfo{booktitle}{Proc.\ TACAS}},
  \bibinfo{series}{LNCS~7795}, \bibinfo{publisher}{Springer}, pp.
  \bibinfo{pages}{641--643}, \doi{10.1007/978-3-642-36742-7_53}.

\bibitemdeclare{inproceedings}{hojjat2018eldarica}
\bibitem{hojjat2018eldarica}
\bibinfo{author}{Hossein \surnamestart Hojjat\surnameend} \&
  \bibinfo{author}{Philipp \surnamestart Rümmer\surnameend}
  (\bibinfo{year}{2018}): \emph{\bibinfo{title}{The ELDARICA Horn Solver}}.
\newblock In: {\slshape \bibinfo{booktitle}{2018 Formal Methods in Computer
  Aided Design (FMCAD)}}, pp. \bibinfo{pages}{1--7},
  \doi{10.23919/FMCAD.2018.8603013}.

\bibitemdeclare{inproceedings}{Hu2020}
\bibitem{Hu2020}
\bibinfo{author}{Qinheping \surnamestart Hu\surnameend}, \bibinfo{author}{John
  \surnamestart Cyphert\surnameend}, \bibinfo{author}{Loris \surnamestart
  D'Antoni\surnameend} \& \bibinfo{author}{Thomas~W. \surnamestart
  Reps\surnameend} (\bibinfo{year}{2020}): \emph{\bibinfo{title}{{Exact and
  approximate methods for proving unrealizability of syntax-guided synthesis
  problems}}}.
\newblock In \bibinfo{editor}{Alastair~F. \surnamestart Donaldson\surnameend}
  \& \bibinfo{editor}{Emina \surnamestart Torlak\surnameend}, editors:
  {\slshape \bibinfo{booktitle}{Proceedings of the 41st {ACM} {SIGPLAN}
  International Conference on Programming Language Design and Implementation,
  {PLDI} 2020, London, UK, June 15-20, 2020}}, \bibinfo{publisher}{ACM}, pp.
  \bibinfo{pages}{1128--1142}, \doi{10.1145/3385412.3385979}.

\bibitemdeclare{inproceedings}{SYMBIOTIC-SVCOMP24}
\bibitem{SYMBIOTIC-SVCOMP24}
\bibinfo{author}{M.~\surnamestart Jonáš\surnameend},
  \bibinfo{author}{K.~\surnamestart Kumor\surnameend},
  \bibinfo{author}{J.~\surnamestart Novák\surnameend},
  \bibinfo{author}{J.~\surnamestart Sedláček\surnameend},
  \bibinfo{author}{M.~\surnamestart Trtík\surnameend},
  \bibinfo{author}{L.~\surnamestart Zaoral\surnameend},
  \bibinfo{author}{P.~\surnamestart Ayaziová\surnameend} \&
  \bibinfo{author}{J.~\surnamestart Strejček\surnameend}
  (\bibinfo{year}{2024}): \emph{\bibinfo{title}{\textsc{Symbiotic 10}: {Lazy}
  Memory Initialization and Compact Symbolic Execution (Competition
  Contribution)}}.
\newblock In: {\slshape \bibinfo{booktitle}{Proc.\ TACAS~(3)}},
  \bibinfo{series}{LNCS~14572}, \bibinfo{publisher}{Springer}, pp.
  \bibinfo{pages}{406--411}, \doi{10.1007/978-3-031-57256-2_29}.

\bibitemdeclare{article}{Kim2021}
\bibitem{Kim2021}
\bibinfo{author}{Jinwoo \surnamestart Kim\surnameend},
  \bibinfo{author}{Qinheping \surnamestart Hu\surnameend},
  \bibinfo{author}{Loris \surnamestart D'Antoni\surnameend} \&
  \bibinfo{author}{Thomas~W. \surnamestart Reps\surnameend}
  (\bibinfo{year}{2021}): \emph{\bibinfo{title}{{Semantics-guided synthesis}}}.
\newblock {\slshape \bibinfo{journal}{Proc. ACM Program. Lang.}}
  \bibinfo{volume}{5}(\bibinfo{number}{POPL}), pp. \bibinfo{pages}{1--32},
  \doi{10.1145/3434311}.

\bibitemdeclare{article}{komuravelli2016smt}
\bibitem{komuravelli2016smt}
\bibinfo{author}{Anvesh \surnamestart Komuravelli\surnameend},
  \bibinfo{author}{Arie \surnamestart Gurfinkel\surnameend} \&
  \bibinfo{author}{Sagar \surnamestart Chaki\surnameend}
  (\bibinfo{year}{2016}): \emph{\bibinfo{title}{SMT-based model checking for
  recursive programs}}.
\newblock {\slshape \bibinfo{journal}{Form. Methods Syst. Des.}}
  \bibinfo{volume}{48}(\bibinfo{number}{3}), p. \bibinfo{pages}{175–205},
  \doi{10.1007/s10703-016-0249-4}.

\bibitemdeclare{inproceedings}{KOJAK-SVCOMP15}
\bibitem{KOJAK-SVCOMP15}
\bibinfo{author}{A.~\surnamestart Nutz\surnameend},
  \bibinfo{author}{D.~\surnamestart Dietsch\surnameend}, \bibinfo{author}{M.~M.
  \surnamestart Mohamed\surnameend} \& \bibinfo{author}{A.~\surnamestart
  Podelski\surnameend} (\bibinfo{year}{2015}):
  \emph{\bibinfo{title}{\textsc{Ultimate Kojak} with Memory Safety Checks
  (Competition Contribution)}}.
\newblock In: {\slshape \bibinfo{booktitle}{Proc.\ TACAS}},
  \bibinfo{series}{LNCS~9035}, \bibinfo{publisher}{Springer}, pp.
  \bibinfo{pages}{458--460}, \doi{10.1007/978-3-662-46681-0_44}.

\bibitemdeclare{article}{somorjai2024bottoms}
\bibitem{somorjai2024bottoms}
\bibinfo{author}{M{\'a}rk \surnamestart Somorjai\surnameend},
  \bibinfo{author}{Mih{\'a}ly \surnamestart Dobos-Kov{\'a}cs\surnameend},
  \bibinfo{author}{Zs{\'o}fia \surnamestart {\'A}d{\'a}m\surnameend},
  \bibinfo{author}{Levente \surnamestart Bajczi\surnameend} \&
  \bibinfo{author}{Andr{\'a}s \surnamestart V{\"o}r{\"o}s\surnameend}
  (\bibinfo{year}{2024}): \emph{\bibinfo{title}{Bottoms up for CHCs: novel
  transformation of linear constrained horn clauses to software verification}}.
\newblock {\slshape \bibinfo{journal}{arXiv preprint arXiv:2404.15215}},
  \doi{10.4204/eptcs.402.11}.

\bibitemdeclare{inproceedings}{THETA-SVCOMP25}
\bibitem{THETA-SVCOMP25}
\bibinfo{author}{C.~\surnamestart Telbisz\surnameend},
  \bibinfo{author}{L.~\surnamestart Bajczi\surnameend},
  \bibinfo{author}{D.~\surnamestart Szekeres\surnameend} \&
  \bibinfo{author}{A.~\surnamestart Vörös\surnameend} (\bibinfo{year}{2025}):
  \emph{\bibinfo{title}{\textsc{Theta}: {Various} Approaches for Concurrent
  Program Verification (Competition Contribution)}}.
\newblock In: {\slshape \bibinfo{booktitle}{Proc.\ TACAS~(3)}},
  \bibinfo{series}{LNCS~15698}, \bibinfo{publisher}{Springer}, pp.
  \bibinfo{pages}{260--265}, \doi{10.1007/978-3-031-90660-2_22}.

\bibitemdeclare{inproceedings}{theta-fmcad2017}
\bibitem{theta-fmcad2017}
\bibinfo{author}{Tam\'as \surnamestart T\'oth\surnameend},
  \bibinfo{author}{\'{A}kos \surnamestart Hajdu\surnameend},
  \bibinfo{author}{Andr\'as \surnamestart V\"or\"os\surnameend},
  \bibinfo{author}{Zolt\'an \surnamestart Micskei\surnameend} \&
  \bibinfo{author}{Istv\'an \surnamestart Majzik\surnameend}
  (\bibinfo{year}{2017}): \emph{\bibinfo{title}{Theta: a Framework for
  Abstraction Refinement-Based Model Checking}}.
\newblock In \bibinfo{editor}{Daryl \surnamestart Stewart\surnameend} \&
  \bibinfo{editor}{Georg \surnamestart Weissenbacher\surnameend}, editors:
  {\slshape \bibinfo{booktitle}{Proceedings of the 17th Conference on Formal
  Methods in Computer-Aided Design}}, pp. \bibinfo{pages}{176--179},
  \doi{10.23919/FMCAD.2017.8102257}.

\bibitemdeclare{inproceedings}{Wesley2022}
\bibitem{Wesley2022}
\bibinfo{author}{Scott \surnamestart Wesley\surnameend}, \bibinfo{author}{Maria
  \surnamestart Christakis\surnameend}, \bibinfo{author}{Jorge~A. \surnamestart
  Navas\surnameend}, \bibinfo{author}{Richard~J. \surnamestart
  Trefler\surnameend}, \bibinfo{author}{Valentin \surnamestart
  W{\"{u}}stholz\surnameend} \& \bibinfo{author}{Arie \surnamestart
  Gurfinkel\surnameend} (\bibinfo{year}{2022}):
  \emph{\bibinfo{title}{{Verifying Solidity Smart Contracts via Communication
  Abstraction in SmartACE}}}.
\newblock In \bibinfo{editor}{Bernd \surnamestart Finkbeiner\surnameend} \&
  \bibinfo{editor}{Thomas \surnamestart Wies\surnameend}, editors: {\slshape
  \bibinfo{booktitle}{Verification, Model Checking, and Abstract Interpretation
  - 23rd International Conference, {VMCAI} 2022, Philadelphia, PA, USA, January
  16-18, 2022, Proceedings}}, {\slshape \bibinfo{series}{Lecture Notes in
  Computer Science}} \bibinfo{volume}{13182}, \bibinfo{publisher}{Springer},
  pp. \bibinfo{pages}{425--449}, \doi{10.1007/978-3-030-94583-1\_21}.

\bibitemdeclare{inproceedings}{ESBMC-SVCOMP25}
\bibitem{ESBMC-SVCOMP25}
\bibinfo{author}{T.~\surnamestart Wu\surnameend},
  \bibinfo{author}{X.~\surnamestart Li\surnameend},
  \bibinfo{author}{E.~\surnamestart Manino\surnameend},
  \bibinfo{author}{R.~\surnamestart Menezes\surnameend},
  \bibinfo{author}{M.~\surnamestart Gadelha\surnameend},
  \bibinfo{author}{S.~\surnamestart Xiong\surnameend},
  \bibinfo{author}{N.~\surnamestart Tihanyi\surnameend},
  \bibinfo{author}{P.~\surnamestart Petoumenos\surnameend} \&
  \bibinfo{author}{L.~\surnamestart Cordeiro\surnameend}
  (\bibinfo{year}{2025}): \emph{\bibinfo{title}{\textsc{ESBMC v7.7}:
  {Efficient} Concurrent Software Verification with Scheduling, Incremental
  {SMT} and Partial Order Reduction (Competition Contribution)}}.
\newblock In: {\slshape \bibinfo{booktitle}{Proc.\ TACAS~(3)}},
  \bibinfo{series}{LNCS~15698}, \bibinfo{publisher}{Springer}, pp.
  \bibinfo{pages}{223--228}, \doi{10.1007/978-3-031-90660-2_16}.

\end{thebibliography}

\end{document}